\documentclass[12pt]{article}
\def\slash#1{\setbox0=\hbox{$#1$}#1\hskip-\wd0\hbox to\wd0{\hss\sl/\/\hss}}
\usepackage{epsf, cite, amsmath, amssymb}
\usepackage{epsfig}
\usepackage[all]{xy}
\makeatletter
\renewcommand\section{\@startsection {section}{1}{\z@}%
                                   {-3.5ex \@plus -1ex \@minus -.2ex}
                                   {2.3ex \@plus.2ex}%
                                   {\normalfont\large\bfseries}}
\renewcommand\subsection{\@startsection{subsection}{2}{\z@}%
                                     {-3.25ex\@plus -1ex \@minus -.2ex}%
                                     {1.5ex \@plus .2ex}%
                                     {\normalfont\bfseries}}
\makeatother

\let\non\nonumber

\newcommand{\bea}{\begin{eqnarray}}
\newcommand{\eea}{\end{eqnarray}}
\newcommand{\be}{\begin{equation}}
\newcommand{\ee}{\end{equation}}

\newcommand{\Z}{{\mathbb Z}}

\newcommand{\p}{\partial}

\newcommand{\del}{\delta}
\newcommand{\Ot}{{\mathcal{O}}_\tau}

\newcommand{\OT}{{\mathcal{O}}_2}
\newcommand{\Om}{\mathcal{O}}

\newcommand{\dal}{\dot\alpha}

\newcommand{\C}[1]{$(\ref{#1})$}

\newcommand{\bL}{\Lambda}
\newcommand{\ap}{\alpha'}

\begin{document}
\begin{titlepage}

\begin{center}

{June 29, 2004}
\hfill                  hep-th/0406267

\hfill DAMTP-2004-70 , EFI-04-21

\vskip 2 cm
{\Large \bf A Curious Truncation of N=4 Yang-Mills}\\
\vskip 1.25 cm { Anirban Basu$^{a}$\footnote{email: basu@theory.uchicago.edu},
Michael
 B. Green$^{b}$\footnote{email: M.B.Green@damtp.cam.ac.uk} and Savdeep
  Sethi$^{a}$\footnote{email: sethi@theory.uchicago.edu}
}\\
{\vskip 0.75cm
$^a$ Enrico Fermi Institute, University of Chicago,
Chicago, IL 60637, USA\\
\vskip 0.2cm
$^b$ Department of Applied Mathematics and Theoretical Physics,
Wilberforce Road, Cambridge,
CB3 OWA, UK\\
}

\end{center}

\vskip 2 cm

\begin{abstract}
\baselineskip=18pt

The coupling constant dependence of correlation functions  of BPS operators
in N=4 Yang-Mills can be expressed in terms of  integrated
correlation functions. 
We approximate these integrated correlators by
using a truncated OPE expansion.
This leads to differential equations for the coupling
dependence. When applied to a particular sixteen point correlator, 
the coupling dependence we find agrees with the corresponding
amplitude computed via the AdS/CFT correspondence. We conjecture  that this
truncation 
becomes exact in the large $N$ and large 't Hooft coupling limit.

\end{abstract}

\end{titlepage}

\pagestyle{plain}
\baselineskip=18pt


Among the consequences of space-time supersymmetry in type IIB string
theory is the determination of certain higher derivative
interactions. One example is the $16$ dilatino
interaction
\be \label{f16}
\frac{1}{\ap} \int d^{10} x {\sqrt{-g}} e^{-\phi /2}
f^{(12,-12)} (\tau,\bar\tau) \Lambda^{16} (x)\, ,\ee
where $\phi$ is the dilaton and $\Lambda$ is the dilatino, which is a
sixteen-component chiral spinor.  The $SO(9,1)$ Lorentz structure appearing in~\C{f16}\ is
the unique singlet in the product of $16$ dilatinos.
This interaction is one of the leading corrections to
the type IIB supergravity action in an $\alpha'$ expansion.
The dependence on the string
coupling, $\tau$, is encoded in the modular form  $f^{(12,-12)}
(\tau,\bar\tau)$ which has weights $(12,-12)$. This
coupling dependence was conjectured in~\cite{Green:1998me,
  Green:1997as} and shown to be a consequence of space-time
supersymmetry in~\cite{Green:1998by}. Under an $SL(2,\Z)$
transformation,
\be
\tau \rightarrow  \left( \frac{a\tau+b}{c\tau+d}\right),
\qquad a,b,c,d\in \Z \, ,
\ee
a modular form with weights $(w, \bar{w})$ transforms in the following way:
\be f^{(w,\bar{w})} (\tau,\bar\tau) \rightarrow (c\tau +d)^w (c\bar\tau
+d)^{\bar{w}} f^{(w,\bar{w})} (\tau,\bar\tau)\, .\ee
{}For a detailed review of these issues, see~\cite{Green:1999qt}.

Via the AdS/CFT correspondence~\cite{Maldacena:1998re}, we are led to
consider a particular correlator of sixteen operators in N=4 Yang-Mills which
can naturally detect the chiral space-time interaction~\C{f16}.  The
operator dual to the dilatino is denoted by $\bL^i_\alpha(x)$, where
$\alpha=1,2$ and $i=1,\ldots 4$. In N=4
Yang-Mills, this operator is special because it sits in the current
multiplet~\cite{Bergshoeff:1981is} and, in particular,
 it is BPS. It transforms in the $({\bf 2}, {\bf 1}, {\bf
  4})$ representation of the $SU(2)_L \times SU(2)_R \times SU(4)_R$
symmetry group.

The N=4 Yang-Mills
correlator we wish to study is,
\be\label{corr}
\langle \bL^{i_1}_{\alpha_1}(x_1) \, \cdots\,
\bL^{i_{16}}_{\alpha_{16}}(x_{16}) \rangle \,.
\ee
{}From the string theory perspective, this correlator should be
renormalized by~\C{f16}\ among other interactions. The instanton contributions to~\C{corr}\
have been investigated in a semiclassical approximation. The
one-instanton contribution was computed for gauge group
$SU(2)$ in~\cite{Bianchi:1998nk} and for $SU(N)$
in~\cite{Dorey:1998xe}. These results were extended to
multi-instantons in the large $N$ limit
in~\cite{Dorey:1999pd}. Finally, the one-instanton analysis has been
extended to more general correlators in~\cite{Green:2002vf}. What is most
remarkable about these results is that they confirm our
expectations from string theory using~\C{f16}; for example, see~\cite{Banks:1998nr}.

Our aim in this letter is to find a field theoretic analogue of the
argument given in~\cite{Green:1998by}. We will make use of recent
results on the coupling dependence in N=4
Yang-Mills~\cite{hasappeared} and  will find an intriguing result.
The argument is applicable to any correlator of
BPS operators. Here, in order to compare with the results
of~\cite{Green:1998by},  we will consider the particular
example of the $\bL^{16}$ correlator~\C{corr}. Let us take the derivative with respect to the
complex coupling~\cite{Intriligator:1998ig}
\be \label{der1}
\frac{\p}{\p\bar\tau}\langle  \prod_{r=1}^{16}
\Lambda^{i_r}_{\alpha_r} (x_r)   \rangle = \langle \frac{\p}{\p\bar\tau} \left( \prod_{r=1}^{16}
\Lambda^{i_r}_{\alpha_r} (x_r) \right)\rangle - \frac{i}{4\tau_2} \int d^4 z \langle \bar\Ot (z)
\prod_{r=1}^{16}  \Lambda^{i_r}_{\alpha_r} (x_r) \rangle\, .
\ee
In the conventions of~\cite{hasappeared}, the Yang-Mills
action is given by,
\be\label{action} S = \frac{i}{4\tau_2} \int d^4 z \left\{  \tau \Ot (z) - \bar\tau
{\bar\Om}_{\tau}(z)\right\}\, , \ee
where $\tau = \theta_{\rm YM}/2\pi + 4\pi i/g_{\rm YM}^2$.

The integrated OPE between $\bar{\Ot}$ and $\bL(x)$ takes the 
form~\cite{hasappeared} \be \label{iOPE} \int d^4z \bar\Ot(z) \bL(x) 
\sim \sum_{i} \Om_i'(x) + \ldots\, , \ee where each $\Om_i'$ is a 
BPS operator in the current multiplet and the omitted terms involve 
long and semi-short operators. The coefficient of each $\Om_i'$ 
in~\C{iOPE}\ is coupling independent. The semi-short operators do 
not develop anomalous dimensions~~\cite{Arutyunov:2000ku, 
Arutyunov:2000im, Eden:2001ec, Dolan:2001tt, Bianchi:2001cm, 
Penati:2001sv, Bianchi:2002rw, Dolan:2002zh}, and at least some of 
these operators correspond to multi-particle states in supergravity. 
Our curious truncation is quite simply to neglect the long operators 
appearing in~\C{iOPE}, which leads to a simple algebraic structure. 
In other words, we just retain all short and semi-short operators.  
As we will discuss later, there is strong motivation for this 
approximation from the AdS/CFT correspondence.

In principle, we should determine which semi-short operators appear 
in~\C{iOPE}.  For example, there is at least one semi-short operator 
appearing in~\C{iOPE}. This operator is related by supersymmetry to 
the semi-short operator (transforming in the ${\bf 20}$ of 
$SU(4)_R$) which appears in the $\OT(z) \OT(x)$ 
OPE~\cite{Arutyunov:2000ku}. However, for the particular correlators 
under consideration here, we will not need to explicitly determine 
the semi-short contribution.

So we will proceed by substituting~\C{iOPE}\
into~\C{der1}\ as an approximation to the integrated term and evaluate the right hand side.
We can determine which $\Om_i'$ appear in~\C{iOPE}
by an exact tree-level computation using the
propagators and  explicit operators given in~\cite{hasappeared}.
We learn that
\be \label{OPEOTmore} \int d^4z \bar\Ot (z) \bL^i_\alpha (x)
\sim \bL^i{}_\alpha (x) + \sigma^\mu_{\alpha\dal} \bar{J}_\mu^{i\dal} (x)\, ,
\ee
where the operator $\bar{J}$ is the supercurrent dual to the gravitino.
The constants of proportionality are independent of the coupling, but
scheme-dependent at tree-level~\cite{Penati:2000zv}.
We will, however, determine the first of these constants later by demanding consistency with
$SL(2,\Z)$.
 Using the operator
normalizations described in~\cite{hasappeared}, we note that
\be \frac{\p}{\p\bar\tau}\Lambda  = {i\over 2\tau_2}\Lambda. \ee
Collecting factors, we see that~\C{der1}\ becomes
\be \label{compOPE} \left( \tau_2 \frac{\partial}{\partial \bar\tau}
+iA \right) \langle \prod_{r=1}^{16}
\Lambda^{i_r}_{\alpha_r} (x_r) \rangle
= B \sum_{j=1}^{16}
(\sigma^\mu)_{\alpha_j \dal}
\langle \Lambda^{i_1}_{\alpha_1} (x_1) \cdots
\bar{J}^{i_j \dal}_\mu (x_j)\cdots
\Lambda^{i_{16}}_{\alpha_{16}} (x_{16}) \rangle \,, 
\ee
where $A$ and $B$ are constants. We might worry that the right hand
side of~\C{compOPE}\ is always proportional to a sum of contact terms using
the Ward identity for the spin $1/2$  anomaly
described in~\cite{hasappeared}. This raises a subtle
issue. Indeed, if all the supercurrents are preserved by the vacuum, which is
the case for the topologically
trivial vacuum, then the right hand side  of~\C{compOPE}\ is purely a
sum of contact terms. In an instanton background, this is no longer
the case since half of the supersymmetries are broken. When those
broken currents appear in~\C{compOPE}, the result need not be purely a
sum of contact terms since there is no associated Ward identity.

We will now explain how equation~\C{compOPE}\ meshes with our
expectations from  instanton analysis. The
semi-classical one-instanton contribution to~\C{corr}\ has the
coupling dependence,
\be \label{semi}
(\tau_2)^{12} e^{2\pi i \tau}\, .
\ee
On the other hand, the leading approximation to the right hand side
of~\C{compOPE}\ is suppressed by an extra factor of $g^2_{\rm YM}.$ 
{}For equation~\C{compOPE}\ to make sense, the left hand side
must vanish at leading order. We will later show that $SL(2,\Z)$ covariance
requires $A=-6$. With this value, the left hand side
of~\C{compOPE}\ does indeed vanish using the semi-classical
result~\C{semi}.  In principle, there are infinitely
many checks of~\C{compOPE}\ that involve loop corrections to the
semi-classical instanton result.

Let us examine one term from the right hand side of~\C{compOPE}.
Differentiating with respect to $\tau$ gives,
\bea \label{OPEJL} \frac{\partial}{\partial \tau}
\langle \bar{J}^{i_1 \dal}_\mu (x_1) \prod_{r=2}^{16}
\Lambda^{i_r}_{\alpha_r} (x_r) \rangle
&=& -\frac{8i}{\tau_2} \langle \bar{J}^{i_1 \dal}_\mu (x_1) \prod_{r=2}^{16}
\Lambda^{i_r}_{\alpha_r} (x_r) \rangle
\nonumber \\ &&
+\frac{i}{4\tau_2}  \int d^4 z
\langle \Ot (z) \bar{J}^{i_1 \dal}_\mu (x_1)
\prod_{r=2}^{16} \Lambda^{i_r}_{\alpha_r} (x_r) \rangle\, ,
 \eea
where we have used
\be \frac{\p}{\p\tau} \bar{J}  = - {i\over 2\tau_2}\bar{J}\,, \qquad
\frac{\p}{\p\tau} \bL  = - {i\over 2\tau_2}\bL\, .
\ee
If we are fortunate, we might hope to obtain a pair of coupled
differential equations relating just two correlators.
To see whether this is the case, we first need to
evaluate the integrated OPE between $\Ot$ and $\bL$. The operators are
given in~\cite{hasappeared} and a tree-level computation gives
\be
 \label{OLope} \int d^4z \Ot (z) \bL^i_\alpha (x) \sim \bL^i{}_\alpha (x)\, ,
\ee
where we again neglect long and semi-short operators. The analogous result for
$\bar{J}$ gives
\be \label{OJope}\int d^4z \Ot (z) \bar{J}^{i\dal}_\mu (x) \sim
\bar{J}^{i\dal}_\mu (x) + (\bar\sigma_\mu)^{\dal\alpha} \bL^i{}_\alpha (x)\, .
\ee
On substituting~\C{OLope}\ and~\C{OJope}\ into~\C{OPEJL}, we find that
\be \label{compbarJ} \left(\tau_2 \frac{\partial}{\partial \tau} +iC\right)
\langle \bar{J}^{i_1 \dal}_\mu (x_1) \prod_{r=2}^{16}
\Lambda^{i_r}_{\alpha_r} (x_r) \rangle
=D (\bar\sigma^\mu)^{\dal \alpha_1} \langle \prod_{r=1}^{16}
\bL^{i_r}_{\alpha_r} (x_r) \rangle\, ,
\ee
where $C$ and $D$ are constants.

Now~\C{compOPE}\ and~\C{compbarJ}\ form a system of coupled
differential equations involving the $\bL^{16}$ and $\bar{J} \bL^{15}$
correlators. Inserting~\C{compbarJ} into~\C{compOPE}\ gives an eigenvalue
equation for the $\bL^{16}$ correlator,
\be \label{laplace}
 \left(\tau_2 \frac{\partial}{\partial \tau} +iC\right)
 \left( \tau_2 \frac{\partial}{\partial \bar\tau}
+iA \right) \langle \prod_{r=1}^{16} \Lambda^{i_r}_{\alpha_r} (x_r) \rangle
=E \langle \prod_{r=1}^{16}
\Lambda^{i_r}_{\alpha_r} (x_r) \rangle\, ,
\ee
where $E$ is some constant. This argument, which neglects long and
semi-short 
operators and uses the OPE expansion in the integrated correlators,
gives a result completely analogous to the supergravity analysis
of~\cite{Green:1998by}!

Now we will determine $A$ and $C$ under the assumption that $SL(2,\Z)$
is an exact symmetry. Note that under an $SL(2,\Z)$ transformation,
\be \frac{\del\tau}{\tau_2} \rightarrow \left(\frac{c\bar\tau
  +d}{c\tau +d} \right)
\frac{\del\tau}{\tau_2}\, .
\ee
Using the invariance of $\del S$, we conclude that $\Ot$
has weights $(1,-1)$ while $\bar\Ot$ has weights $(-1,1)$. The current
multiplet, in which $\Ot$ sits, is generated by the action of
supercharges on the superconformal primary, $\Om_2$. We denote the
supercharges by $\del$ and $\bar\del$ following~\cite{Intriligator:1998ig}.
The operators $\OT, \, \del$ and $\bar\del$ will be assigned modular weights $(p,q), (k,l)$ and
$(l,k)$, respectively.   Using the relations $\Ot =\del^4 \OT$, $\bar\Ot =\bar\del^4
\OT$ and the modular weights deduced above  we
see that
\be \label{consrmod} k-l =\frac{1}{2}\,, \qquad p=q \, .\ee
{}From the relation $\{\del,\bar\del\} \sim \partial_\mu$, it follows that
$\del\bar\del$ must have weights $(0,0)$ because taking a space-time
derivative is an operation that commutes
with $SL(2,\Z)$. This leads to the relation,
\be \label{consrmod2} k+l=0 \quad \Rightarrow \quad k=\frac{1}{4}\,, \quad l=-\frac{1}{4}
\, . \ee
Using the defining relation for the $R$-charge, $[\del,R] \sim \del$,
we see that $R$ has weights $(0,0)$. Since $R=\del\bar\del \OT$, we
deduce that $p=q=0$.

Therefore $\bL =\del^3 \OT$ has weights $(3/4,-3/4)$
so that $\bL^{16}$ has weights $(12,-12)$. This is in accord with our
expectations from gravity. We also see that $\bar{J}$ has weights
$(-1/4, 1/4)$ so $\bar{J} \bL^{15}$ has weights
$(11,-11)$. To be consistent with $SL(2,\Z)$, the scheme-dependent
coefficients $A$ and $C$ must take the values
\be A = -6 \quad {\rm and} \quad C = -\frac{11}{2}\, ,\ee
to ensure that~\C{laplace}\ is modular covariant.
{}From these $SL(2,\Z)$ transformation properties,
we can predict some of the tree-level contact terms
between $\Ot$ and $\bL$ or $\bar{J}$
\bea
&&\int d^4z \Ot(z) \bL(x) = \frac{7}{2} \bL(x)\, ,
\qquad  \int d^4z \bar\Ot(z) \bL(x) = \frac{1}{2} \bL(x)+\ldots, \non\\
&&\int d^4z \Ot(z) \bar{J}(x) = \frac{3}{2}\bar{J}(x)+\ldots \, ,
\qquad \int d^4z \bar\Ot(z) \bar{J}(x) = \frac{5}{2}\bar{J}(x)\, .
\eea
The omitted terms are additional current multiplet operators whose coefficients
are undetermined.
Defining modular covariant derivatives
\be D_w = i\left(\tau_2 \frac{\partial}{\partial \tau} -\frac{iw}{2}\right),
\qquad  \bar{D}_{\hat{w}} = -i\left(\tau_2
\frac{\partial}{\partial \bar\tau} +\frac{i\hat{w}}{2}\right)\, , \ee
we see that~\C{laplace}\ becomes
\be \label{L16eqn}
D_{11} \bar{D}_{-12}\langle \prod_{r=1}^{16} \Lambda^{i_r}_{\alpha_r} (x_r)\rangle =
E \langle\prod_{r=1}^{16} \Lambda^{i_r}_{\alpha_r} (x_r) \rangle \, .
\ee
We  will determine the eigenvalue $E$  by making use of known
results in the instanton sector.

To describe the solutions of~\C{L16eqn}, we need to know the domain of
$\tau$. In the full $SL(2,\Z)$ invariant theory, $\tau$ takes values
in the fundamental domain. It is possible, although it appears unlikely,
that our truncation of  the OPE does not respect $SL(2,\Z)$. 
We will assume this is not the case since our
truncation involves no regulator and
because the neglected long operators are likely to give rise to different
space-time structures. The striking agreement between~\C{L16eqn}\ and
the equations from supergravity~\cite{Green:1998by}\ is further evidence for this
assumption.

Assuming power law behavior in $\tau_2$ as $\tau_2\to\infty$,~\C{L16eqn}\ has a unique
solution characterized by $\ell$. The solutions are given by
\be \label{fg} f_\ell^{(12,-12)} (\tau,\bar\tau) =\sum_{(m,n) \neq (0,0)}
\frac{\tau_2^{\ell+ 1/2}}
{(m+n\tau)^{\ell+ 25/2} (m+n\bar\tau)^{\ell- 23/2}}\, ,
\ee
for $\ell \geq 1$. These modular forms satisfy~\C{L16eqn}\ with
eigenvalue
\be
E = \frac{1}{4} (\ell^2 -\frac{529}{4})\, .
\ee
The leading semi-classical $k$-instanton contribution
to~\C{fg}\ has the form,
\be  f_\ell^{(12,-12)}(\tau,\bar\tau) \sim ({\tau_2})^{12} k^{\ell+ 23/2}e^{2\pi i k \tau}
\sum_{m|k} \left(\frac{1}{m}\right)^{2\ell} + \ldots\, . \ee
In order to determine the value of $E$, we will compare this expression with
information from semi-classical $k$-instanton computations in N=4 Yang-Mills.
These were computed in~\cite{Dorey:1999pd}\  in the large $N$ approximation
with the
't Hooft coupling, $\lambda= g^2_{YM} N \rightarrow 0$. It is easy to determine
that ${\ell}=1$, which gives
\be\label{Nev}
E(N\rightarrow \infty) = - \frac{525}{16}\, .
\ee
This answer should also be determinable directly from a one-loop correction to
the one-instanton contribution in the Yang--Mills theory,
which is an interesting computation in its own right.

The instanton computation picks out a particular space-time structure
in~\C{corr}. This structure agrees with the space-time dependence
found in the gravity computation using the coupling~\C{f16}
(see~\cite{Bianchi:1998nk}). The structure is
completely antisymmetric in the sixteen inserted operators. The 
precise space-time  structure, however, is not important for our
discussion. What is unusual about this result  is that
the leading behavior at weak coupling ($\tau_2 \rightarrow \infty$)
is non-analytic in the Yang-Mills coupling
\be\label{perturb}
 f_1^{(12,-12)}(\tau,\bar\tau) \rightarrow (\tau_2)^{3/2} + \ldots\, .
\ee
This behavior cannot be seen in standard perturbation theory at fixed $N$. To
obtain the analytic behavior in the coupling, which we certainly expect
in perturbative Yang-Mills, there must be a series of corrections
to~\C{Nev}\ in powers of $1/N$. These corrections correspond to
additional modular forms beyond $f_1^{(12,-12)}(\tau, \bar\tau)$ whose
sum (should they be summable) might give a
perturbative expansion analytic in the coupling, $g^2_{YM}$.

We also learn from~\C{compOPE}\ that
\be \langle \bar{J} \bL^{15} \rangle \sim f_1^{(11,-11)} (\tau,\bar\tau)
=\sum_{(m,n) \neq (0,0)} \tau_2^{3/2}
\frac{(m+n\bar\tau)^{19/2}}{(m+n\tau)^{25/2}}\, ,\ee
where the space-time dependence is omitted. As before,
equation~\C{compOPE}\ selects the particular space-time structure that
emerges both in gravity and the semi-classical instanton computation.

The agreement of this analysis with the corresponding supergravity 
analysis~\cite{Green:1998by}\ does not require consideration of 
semi-short operators. Any contribution from semi-short operators can 
be absorbed into a redefinition of the parameters $D$ and $E$. 
However, had we considered the $\tau$ rather than $\bar\tau$ 
derivative of~\C{corr}, we would have been forced to consider 
semi-short operators. Using the OPE~\C{OLope}\ which omits 
semi-short operators, we would conclude that \be\label{vanishing} 
D_{12} \langle \Lambda^{16}\rangle = 0. \ee This would contradict 
the result of our analysis. What must correct~\C{vanishing}\ is a 
semi-short contribution, which corresponds to a multi-particle state 
in gravity. Using the results of~\cite{Arutyunov:2000ku}\ extended 
to $\Ot(z) \bL(x)$, we see that there is a semi-short operator in 
the product. This composite operator must be regularized as $z 
\rightarrow x$ along the lines described in~\cite{Bianchi:2003eg}. 
At least heuristically, this gives a contribution with the desired 
structure. 

We are left with the fascinating question as to why our truncation
of the OPE is sensible.   
The agreement   with supergravity
computations suggests that our truncation is valid for large $N$ and large 't
Hooft coupling, $\lambda \rightarrow \infty$. This seems plausible
because the AdS/CFT correspondence teaches us
that the anomalous dimensions of many (and perhaps all) long operators
become large as $\lambda \rightarrow \infty$. 
More precisely,
the OPE expansion is only valid when $\Ot(z)$ approaches
$\Om(x)$ so that $|z-x|$ is small compared to the distance between
$\Om(x)$ and any other inserted operator. In the OPE approximation, a
long operator appears in the form
\be
{\Ot}(z) \Om(x) \sim  |z-x|^{\Delta_L - \Delta -4}
\Om_L(x) + \ldots\, ,
\ee
where the BPS operator $\Om(x)$ has conformal dimension $\Delta$,
while the long operator $\Om_L(x)$ has dimension $\Delta_L$. As
$\lambda \rightarrow \infty$, the contribution from the long operator
is therefore suppressed because $\Delta_L\rightarrow \infty$.
What this does not explain is why the OPE
approximation to the integral in~\C{der1}\ becomes exact
as $\lambda \rightarrow \infty$.    It should be noted that the truncation is 
incompatible with the small 't Hooft coupling limit, $\lambda\to 0$.  This follows 
from the singular perturbative behavior of $f^{(12,-12)}_1$ given in~\C{perturb}. 

This same truncation procedure can be applied to any correlator of
BPS operators.  Starting with an $n$-point function, the procedure
yields another $n$-point function. Because~\C{iOPE}\ always results in
short operators, $\Om'_i$, in the same supermultiplet as
$\Om$~\cite{hasappeared}, we
always find a closed set of coupled equations (generally
more than two equations) by iterating this procedure. If the
semi-short contribution can be controlled or neglected (as in this
case), these equations should encode 
non-perturbative information about the correlators.

This kind of analysis opens up the
possibility that we might be able to learn about gravity from N=4
Yang-Mills, rather than vice-versa. In particular, recent developments
in supersymmetric Yang-Mills~\cite{Sethi:2004np}\ coupled with gravity
computations~\cite{Berkovits:1998ex, Green:1998by, Sinha:2002zr}\
suggest that there exist
determinable interactions analogous to~\C{f16}\ at higher orders in the
$\alpha'$ expansion. We can hope to learn about these interactions via this
kind of analysis.

\section*{Acknowledgements}
The work of A.~B. is supported in part by  NSF Grant No.
PHY-0204608. The work of M.~B.~G. is supported in part by a PPARC
rolling grant.
The work of S.~S. is supported in part by NSF CAREER Grant No.
PHY-0094328 and by the Alfred P. Sloan Foundation.





\providecommand{\href}[2]{#2}\begingroup\raggedright\endgroup

\end{document}